\documentclass{aa}

\usepackage{graphicx}
\usepackage{txfonts}
\usepackage{hyperref}
\usepackage{CJK}
\hypersetup{colorlinks = true, linkcolor=red, citecolor=blue, urlcolor=blue}
\usepackage{natbib}
\begin{document}
\begin{CJK*}{UTF8}{gbsn}

   \title{LAMOST J171013+532646: a detached short-period non-eclipsing hot subdwarf + white dwarf binary}
    \titlerunning{sdB + WD}
    \authorrunning{M. Yang et al.}
          
    \author{Mingkuan Yang (杨明宽)\inst{1,3}
          \and
          Hailong Yuan (袁海龙)\inst{1}
          \and
          Zhongrui Bai (白仲瑞)\inst{1}
          \and
          Zhenwei Li (李振威)\inst{2,4}
          \and
          Yuji He (何玉吉)\inst{1,3}
          \and
          Xin Huang (黄鑫)\inst{1,3}
          \and
          Yiqiao Dong (董义乔)\inst{1}
          \and
          Mengxin Wang (汪梦欣)\inst{1}
          \and
          Xuefei Chen (陈雪飞)\inst{2,4}
          \and
          Junfeng Wang (王俊峰)\inst{5}
          \and
          Yao Cheng (程瑶)\inst{5}
          \and
          Haotong Zhang (张昊彤)\inst{1}
          }

   \institute{Key Laboratory of Optical Astronomy, National Astronomical Observatories, Chinese Academy of Sciences, Beijing 100101, China\\
              \email{htzhang@bao.ac.cn, yuanhl@bao.ac.cn}·
        \and
            Yunnan Observatories, Chinese Academy of Sciences, 650011, China
        \and
            School of Astronomy and Space Science, University of Chinese Academy of Sciences, Beijing 100049, China
        \and
            International Centre of Supernovae, Yunnan Key Laboratory, Kunming, 650216, People’s Republic of China
        \and
            Department of Astronomy, Xiamen University, Xiamen, Fujian 361005, China
             }

   \date{Received xxx xx, 2024; accepted xxx xx, 2024}

% \abstract{}{}{}{}{} 
% 5 {} token are mandatory
 
  \abstract
   {We present an analysis of LAMOST J171013.211+532646.04 (hereafter J1710), a binary system comprising a hot subdwarf B star (sdB) and a white dwarf (WD) companion. Multi-epoch spectroscopy reveals an orbital period of 109.20279 minutes, consistent with TESS and ZTF photometric data, marking it as the sixth detached system known to harbor a WD companion with a period less than two hours. J1710 is remarkably close to Earth, situated at a distance of only \(350.68^{+4.20}_{-4.21} \, \mathrm{pc}\), with a GAIA G-band magnitude of 12.59, rendering it conducive for continuous observations. The spectral temperature is around 25164 K, in agreement with SED fitting results (\(25301^{+839}_{-743} \, \mathrm{K}\)). The TESS light curve displays ellipsoidal variation and Doppler beaming without eclipsing features. Through fitting the TESS light curve using the Wilson-Devinney code, we determined the masses for the sdB (\(M_1 = 0.44^{+0.06}_{-0.07} \, M_{\odot}\)) and the compact object (\(M_2 = 0.54^{+0.10}_{-0.07} \, M_{\odot}\)), with the compact object likely being a WD. Furthermore, MESA models suggest that the sdB, with a helium core mass of 0.431 \(M_{\odot}\) and a hydrogen envelope mass of \(1.3 \times 10^{-3}\, M_{\odot}\), is in the early helium main-sequence phase. The MESA binary evolution shows that the J1710 system is expected to evolve into a double white dwarf system, making it an important source of low-frequency gravitational waves.}
  

   \maketitle
%
%-------------------------------------------------------------------

\section{Introduction}
\label{intro.sec}
Hot subdwarf B stars (sdBs) reside on the extreme horizontal branch in the Hertzsprung-Russell diagram, characterized by extremely thin hydrogen-rich and helium-poor envelope layer undergoing helium core burning. The typical mass for sdBs is \(0.46 \, M_{\odot}\) \citep{1986A&A...155...33H, 2016PASP..128h2001H}. 
Most sdBs are located in compact binaries, featuring orbital periods \(P_{\text{orb}} < 10 \, \text{days}\) \citep{2001MNRAS.326.1391M, 2004Ap&SS.291..321N}.
For sdBs in binary systems, the interaction between the two companion stars is pivotal for sdB formation. Primary formation channels encompass stable Roche lobe overflow (RLOF), common-envelope (CE) ejection, and double helium WDs merging.
Presently, the prevailing belief is that short-period sdB binaries are formed from losing a substantial amount of angular momentum during CE phase, which leads to losses of most envelope mass \citep{2002MNRAS.336..449H, 2003MNRAS.341..669H}.

\citet{2015A&A...576A..44K} analyzed 142 solved sdB binaries from various samples and found that approximately half of them have WD companions.
The majority of known sdB + WD binary systems have relatively long orbital periods, with little or no mass transfer occurring before the sdB evolves into a WD. Currently, only five detached systems with a WD companion are known to have \(P_{\text{orb}} < 2\,\text{hr}\) \citep{2012ApJ...759L..25V, 2017ApJ...835..131K, 2017ApJ...851...28K, 2021NatAs...5.1052P, 2022ApJ...925L..12K}. 
However, for instance, the object OW J074106.0-294811.0 has an orbital period of only 44 minutes, and analysis suggests that its sdB could instead be a helium WD with a mass of \(0.32\,M_{\odot}\) \citep{2017ApJ...851...28K}.
Similar to the CD-\(30^{\circ}1122\) system with a 70-minute orbital period, these compact sdB + WD binaries could be Type Ia supernova progenitors \citep{2012ApJ...759L..25V, 2013A&A...554A..54G}.
The most compact known sdB binary with the sdB filling its Roche lobe is ZTF J213056.71+442046.5, with an orbital period of only 39 minutes \citep{2020ApJ...891...45K}. 
Furthermore, determining the system parameters and evolutionary stages of these systems, involving sdBs and WDs, might reveal previously unknown processes during the CE phase.

The sdB binaries may evolve into double white dwarf (DWD) systems, making them potential gravitational wave (GW) sources for space-based interferometers such as LISA, TianQin, and AMIGO, which are designed to detect milli-Hz, sub-milli-Hz, and even lower frequency signals, and are expected to significantly enhance our ability to detect and study DWD systems, providing valuable insights into the binary evolution process before merger \citep{2018EPJWC.16801004N, 2020PhRvD.102l3547S, 2023LRR....26....2A}.
To date, no GW events from DWD mergers have been detected. However, the LISA mission and Taiji—a planned space-based GW detector with similar sensitivity and capable of operating in tandem with LISA—are expected to detect approximately 10,000 DWD systems, most of which are anticipated to lie within the 3–6 mHz frequency range \citep{2001AcAau..48..549E, 10.1093/nsr/nwx116,2020NatAs...4..108R, CAI20241072, 2024arXiv240207571C}. Detecting GWs from individual systems during their inspiral and merger phases will deepen our understanding of merger processes and provide new opportunities to study the detailed structure of the Milky Way \citep{Jennrich2011}.

Using LAMOST data, \citet{2018ApJ...868...70L} first identified J1710, a system containing an sdB star, and provided its basic atmospheric parameters. 
In this paper, we present a comprehensive analysis of J1710 using optical spectroscopy, Gaia DR3 parallax \citep{2021A&A...649A...1G}, and light curves from TESS \citep{2014SPIE.9143E..20R} and ZTF \citep{2019PASP..131a8002B}, concluding that it consists of an sdB star with a WD companion, with an orbital period of less than two hours.

This paper is organized as follows. 
In Sect. \ref{sec.obs} we compile and provide both existing data and new observations. 
Section \ref{sec:results} describes the methods and results of data analysis. 
In Sect. \ref{sec.mesa}, we present discussions on stellar parameters and binary evolution of J1710.
Finally, conclusions are summarized in Sect. \ref{sec.conclusion}.
For clarity, we designate the visible sdB component, which exhibits orbital radial velocity variations, as the primary (Star 1), and the unseen WD as the secondary (Star 2).
\begin{table*}
%\bc
%\begin{minipage}[]{100mm}
\caption{Spectroscopic observations and estimated parameters of J1710.
\label{tab.sp_obs}}
%\end{minipage}
\setlength{\tabcolsep}{5.0pt}
\begin{center}
\begin{tabular}{cccccccccc}
\hline\noalign{\smallskip}

DATE & BJD & RV & Phase  & $T_{\rm eff}$ & log $g$ & \(\log (n(\text{He})/n(\text{H}))\) & S/N & EXPTIME & Tel.$/$Ins. \\

     & day & km/s  &     & K             & dex     & dex    &     & s & \\
\hline\noalign{\smallskip} 
20130427 & 2456410.2785 &  -66.5$\pm$15.9  & 0.486 &  24610 & 5.57 & -2.15 &  19 &900 & LAMOST$/$LRS \\
20130427 & 2456410.2938 &  176.5$\pm$12.8  & 0.688 &  25186 & 5.76 & -2.10 &  25 &900 & LAMOST$/$LRS \\
20160424 & 2457503.2820 &  -222.6$\pm$7.6  & 0.352 &  25401 &5.61 & -2.30 &  62 &600 & LAMOST$/$LRS \\
20160424 & 2457503.2910 & -75.1$\pm$7.5  & 0.471 & 25137 & 5.55 & -2.32 & 65 &600 & LAMOST$/$LRS \\
20160424 & 2457503.3007 & 91.4$\pm$7.5  & 0.600 & 25811 & 5.72 & -2.32 & 75 &600 & LAMOST$/$LRS \\

20210205 & 2459251.0652 & -83.4$\pm$7.3 & 0.458 & 25387 & 5.62 & -2.19 &  93 &120 & P200$/$DBSP \\
20210206 & 2459252.0534 & -39.4$\pm$7.4  & 0.489 & 25267 & 5.77 & -2.23 &  85 &100 &  P200$/$DBSP \\
20210206 & 2459252.0852 & 84.4$\pm$7.4  & 0.908 & 24942 & 5.85 & -2.27 &  64 &100 &  P200$/$DBSP \\
20210206 & 2459252.0880 & 50.4$\pm$7.6   & 0.945 &  24177 & 5.74 & -2.30 &  44 &100 & P200$/$DBSP \\
20210206 & 2459252.0894 & 15.6$\pm$9.8  & 0.964 & 23653 & 5.64 & -2.32 &  29 &100 & P200$/$DBSP \\

20210603 & 2459368.8871 & -176.5$\pm$7.4   & 0.114 &  25917 & 5.68 & -2.35 &  65 &100 & P200$/$DBSP \\
20210603 & 2459368.8912 & -236.5$\pm$7.4  & 0.167 &  26033 & 5.66 & -2.37 &  67 &100 & P200$/$DBSP \\
20210603 & 2459368.9644 & -204.2$\pm$8.5   & 0.133 &  26148 & 5.70 & -2.34 &  61 &100 & P200$/$DBSP \\
20220223 & 2459634.1005 & -221.7$\pm$2.2   & 0.344 &  24974 & 5.68 & -2.16 &  33 &1100 & CFHT$/$ESPaDOnS \\
20220223 & 2459634.1135 & -23.3$\pm$3.5   & 0.515 &  24825 & 5.63 & -2.23 &  33 &1100 & CFHT$/$ESPaDOnS \\
\noalign{\smallskip}\hline
\end{tabular}
\tablefoot{The DATE column represents the time zone observation night,
and the BJD represents the barycentric time at the middle of exposure.
The RVs has been corrected to the barycenter, and the phase smearing effected is also corrected.}
\end{center}
\smallskip
%{This table is available in its entirety in machine-readable forms in the online journal.}
\end{table*}

\section{Observations and data reduction}
\label{sec.obs}

\subsection{Spectroscopic observations}
\label{sec:spectra}

As of 2023, with the release of LAMOST \citep{1996ApOpt..35.5155W, 2004ChJAA...4....1S} Dr10 data\footnote{http://www.lamost.org/dr10/}, the total number of spectra released has exceeded 20 million.
J1710 was observed by LAMOST on two nights in 2013 and 2016, acquiring six sub-exposure spectra (R $\approx$ 1800) covering the wavelength from 3700 to 9100 \r{A}.
All spectra are processed by LAMOST pipeline, with detailed results presented in Table \ref{tab.sp_obs} \citep{2017PASP..129b4004B, 2021RAA....21..249B}. 
One spectrum from 2013 is excluded due to the low signal-to-noise ratio (S/N = 14) near dawn.
The remaining LAMOST spectra exhibit radial velocity (RV) variations exceeding \(400 \, \mathrm{km/s}\).

Given the significant RV variation observed in the LAMOST spectra, J1710 underwent observations on three nights in 2021 using the Double Spectrograph (DBSP) instrument mounted on the 200-inch Hale Telescope (P200) at Palomar Observatory\footnote{http://info.bao.ac.cn/tap/}. 
Eight individual spectra were acquired, covering a wavelength range of 3900 to 5500 \r{A} in the blue band and 5800 to 7400 \r{A} in the red band, with resolutions of approximately 2100 in the blue band and approximately 3200 in the red band. 
All raw data from the P200 telescope were processed using the IRAF package \citep{1986SPIE..627..733T, 1993ASPC...52..173T}, and RVs are determined utilizing the template cross-correlation method described in \citet{2021RAA....21..249B}. 

On February 23, 2022, CFHT\footnote{https://www.cadc-ccda.hia-iha.nrc-cnrc.gc.ca/en/cfht/} also conducted observations of J1710 and obtained two high-resolution spectra using ESPaDOnS \citep{2003ASPC..307...41D}. ESPaDOnS is an optical spectropolarimeter with a wavelength coverage between 3600 and 10000 \r{A}. The mean spectral resolution of the spectra is approximately 85000, with an exposure time of 1100 s and an average S/N of about 33. Relevant information is summarized in Table \ref{tab.sp_obs}.

% https://www.aanda.org/articles/aa/pdf/2024/04/aa48750-23.pdf

\subsection{Light curves}
\label{sec:lightcurve} 

J1710, as TIC 367779738 in TESS input catalog, has been observed multiple times by TESS \citep{2014SPIE.9143E..20R} from 2019 to 2023 with cadences of 20, 120, 200, and 600 seconds. Given a period of approximately 109 minutes obtained from RV analysis, data with a 120-second cadence for sectors 50 to 53 are downloaded from the MAST archive\footnote{https://mast.stsci.edu}. We used the SAP-FLUX light curves at 120-second cadence due to their lower uncertainty.

We retrieved J1710's light curves from the ZTF\footnote{https://www.ztf.caltech.edu/} \citep{2019PASP..131a8002B} in the g, r, and i bands spanning four years from September 2018 to August 2022. To improve data quality, data points with 'catflags' value of zero and those within one standard deviation of the mean value of the 'chi' flag for each band were retained.

The TESS and ZTF light curves were separately normalized and combined to yield two normalized light curves. The Lomb-Scargle periodogram analysis  \citep{1976Ap&SS..39..447L, 1982ApJ...263..835S} was conducted using the ASTROPY library \citep{2018AJ....156..123A} in Python, identifying an orbital period of P = 109.203(3) minutes. No pulsation features were detected in the power spectrum of the TESS light curve.

%Using the ephemeris established in Sect. \ref{sec:orbit}, the TESS data were phase-folded and re-binned into 400 data points, while the ZTF data were phase-folded in each band. 
%The resulting normalized light curves are displayed in Fig. \ref{fig:RVLC}.

\begin{figure*}[ht]
  \centering
  \includegraphics[width=0.98\textwidth]{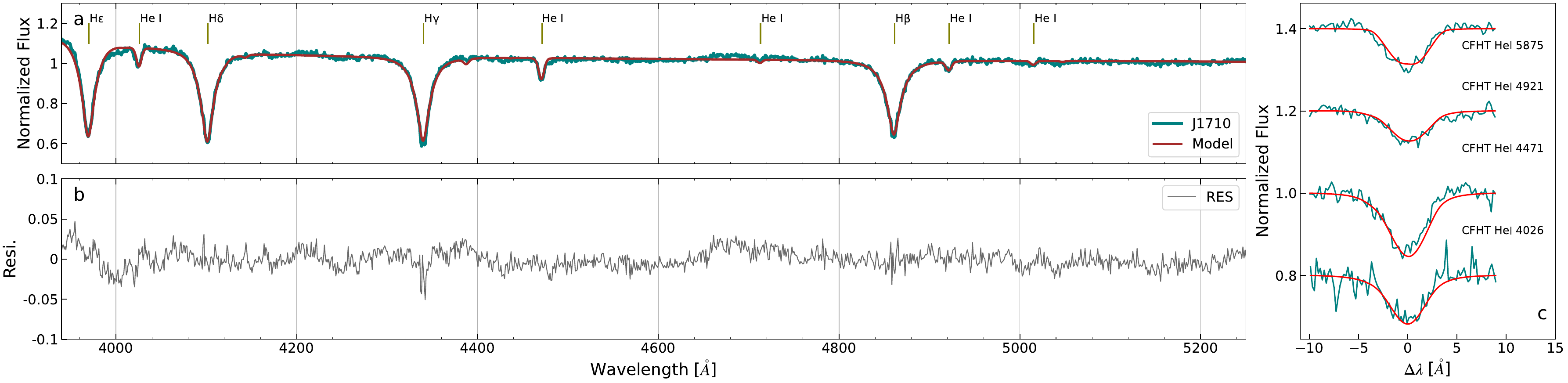} 
  \hfill
  \caption{
Spectrum fitting of J1710. 
Panel a displays the LAMOST combined spectrum in the blue band (green) alongside its best-fitting model (red), with absorption lines of H and He labeled.
Panel b presents the residuals of the fit.
Panel c shows fits of \(v_{\text{rot}}\sin i\) to the helium lines observed in the reduced CFHT spectra, with the normalized fluxes of the single lines shifted for better visualization. 
}
  \label{fig:spec}
\end{figure*}

\section{Methods and results}
\label{sec:results}

\subsection{Spectral parameters}
\label{sec:spparam}

We employed model spectra from previous studies that classified J1710 as an sdB \citep{2007ApJS..169...83L, 2014ASPC..481...95N, 2017arXiv170601859H} and utilized the UlySS package \citep{2009A&A...501.1269K, 2014IAUS..306..340W} to fit the optical spectra of J1710 and derive stellar parameters. 
An example of spectrum fitting is illustrated in panel a of Fig. \ref{fig:spec}.

From panel a of Fig. \ref{fig:spec}, the LAMOST spectrum (green line) exhibits prominent hydrogen Balmer and He\uppercase\expandafter{\romannumeral1} lines, suggesting that the visible star is likely an sdB. 
Upon careful examination of the CFHT high-resolution spectra, the absence of emission line features indicates that the Roche lobe of the visible star is not filled. 
Several weaker lines of C\uppercase\expandafter{\romannumeral2}, N\uppercase\expandafter{\romannumeral2}, O\uppercase\expandafter{\romannumeral2}, Mg\uppercase\expandafter{\romannumeral2}, Al\uppercase\expandafter{\romannumeral3}, Si\uppercase\expandafter{\romannumeral3}, S\uppercase\expandafter{\romannumeral2}, Ti\uppercase\expandafter{\romannumeral3}, V\uppercase\expandafter{\romannumeral3}, and Fe\uppercase\expandafter{\romannumeral3} can be carefully identified. 
A more detailed analysis of metallicity is beyond the scope of this work and will not be further elaborated upon here.
% However, due to the low S/N, a detailed discussion of metal abundances is not included.

The best-fitting parameters for each observed spectrum were determined by minimizing the $\chi^2$ value and are listed in Table \ref{tab.sp_obs}. The mean and standard deviation of these parameters are calculated and adopted as the final values: $T_{\rm eff}=25164\pm663 \, \mathrm{K}$, $\log g = 5.68 \pm 0.08$ dex, and \(\log (n(\text{He})/n(\text{H})) = -2.26 \pm 0.09\) dex. 
The CFHT spectra had an exposure time of 1100 s, accounting for 17\% of the period, and exhibited a low S/N, significantly impacting the broadening of absorption lines.
Therefore, Gaussian smoothing was applied to the CFHT spectra to reduce the resolution to 25000. 
Additionally, based on the RV curve fitted in Sect. \ref{sec:orbit}, corrections for long exposures were applied to the CFHT spectra. The RV at each second of the exposure was calculated and added to the model spectra. These velocity-shifted spectra were then summed and normalized before being compared to the observed spectra.
From the four He\uppercase\expandafter{\romannumeral1} lines (4026, 4471, 4921, 5875 \r{A}), the projected rotational velocity \(v_{\text{rot}} \sin i\) was determined to be \(89\pm12 \, \mathrm{km/s}\), as shown in panel c of Fig. \ref{fig:spec} and summarized in Table \ref{tab.wdfit}.

\begin{figure}[htbp!]
\center
\includegraphics[width=0.48\textwidth]{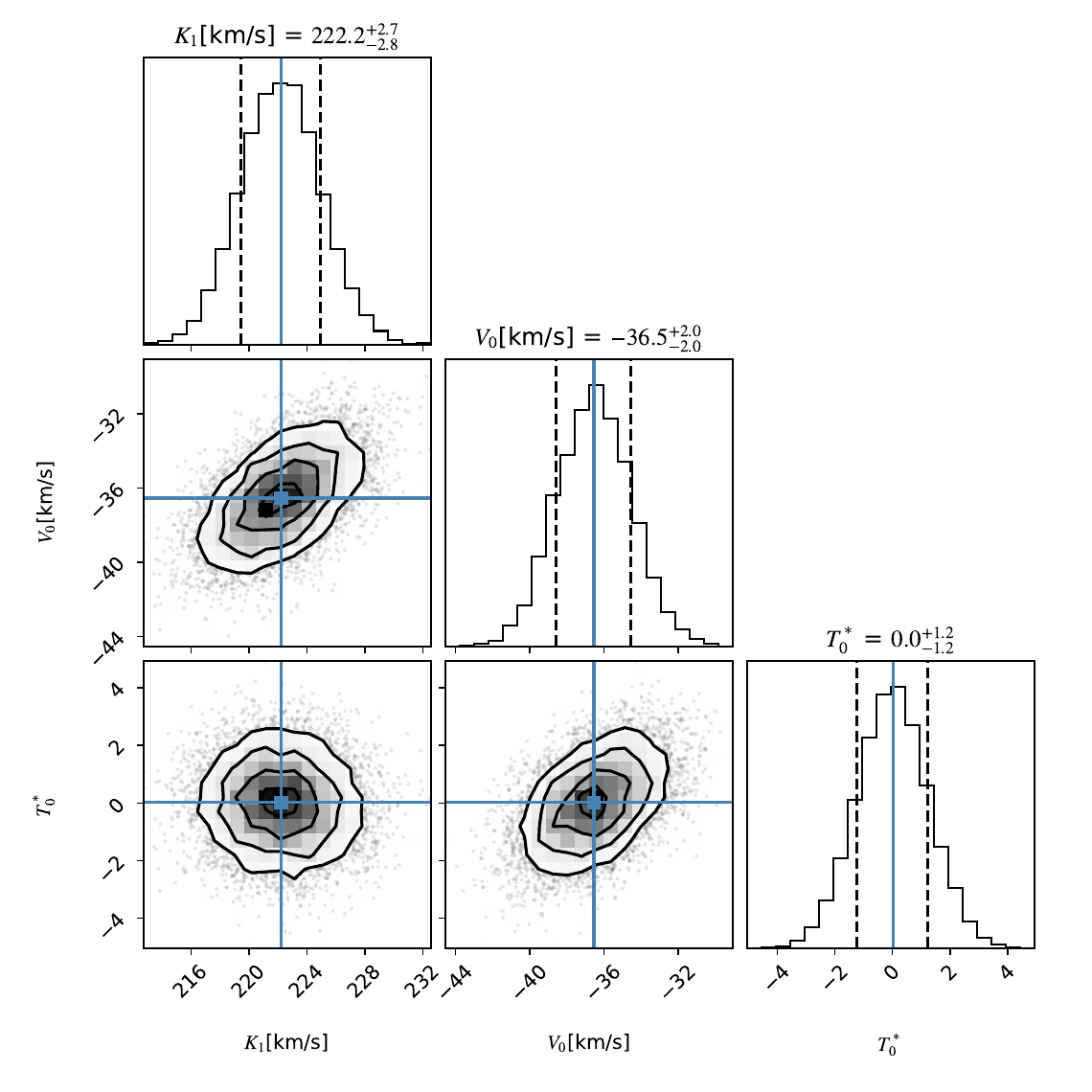}
\caption{Results of sinusoidal RV fitting using MCMC technique with
the eccentricity \(e\) fixed to 0. 
For convenience, \(T_0^*\) is defined as \((T_0 - 2459744.03573) \times 10000\). The derived reference ephemeris is \(T_0 = 2459744.03573 \pm 0.00012\) days, with an RV semi-amplitude of \(K_1 = 222.2^{+2.7}_{-2.8}\) km/s and a systemic velocity of \(V_0 = -36.5 \pm 2.0\) km/s.
}
\label{fig:rvmcmc}
\end{figure}

\begin{figure*}[ht]
  \begin{minipage}[b]{0.48\textwidth} % 创建一个minipage，宽度为页面宽度的一半
    \centering
    \includegraphics[width=\textwidth]{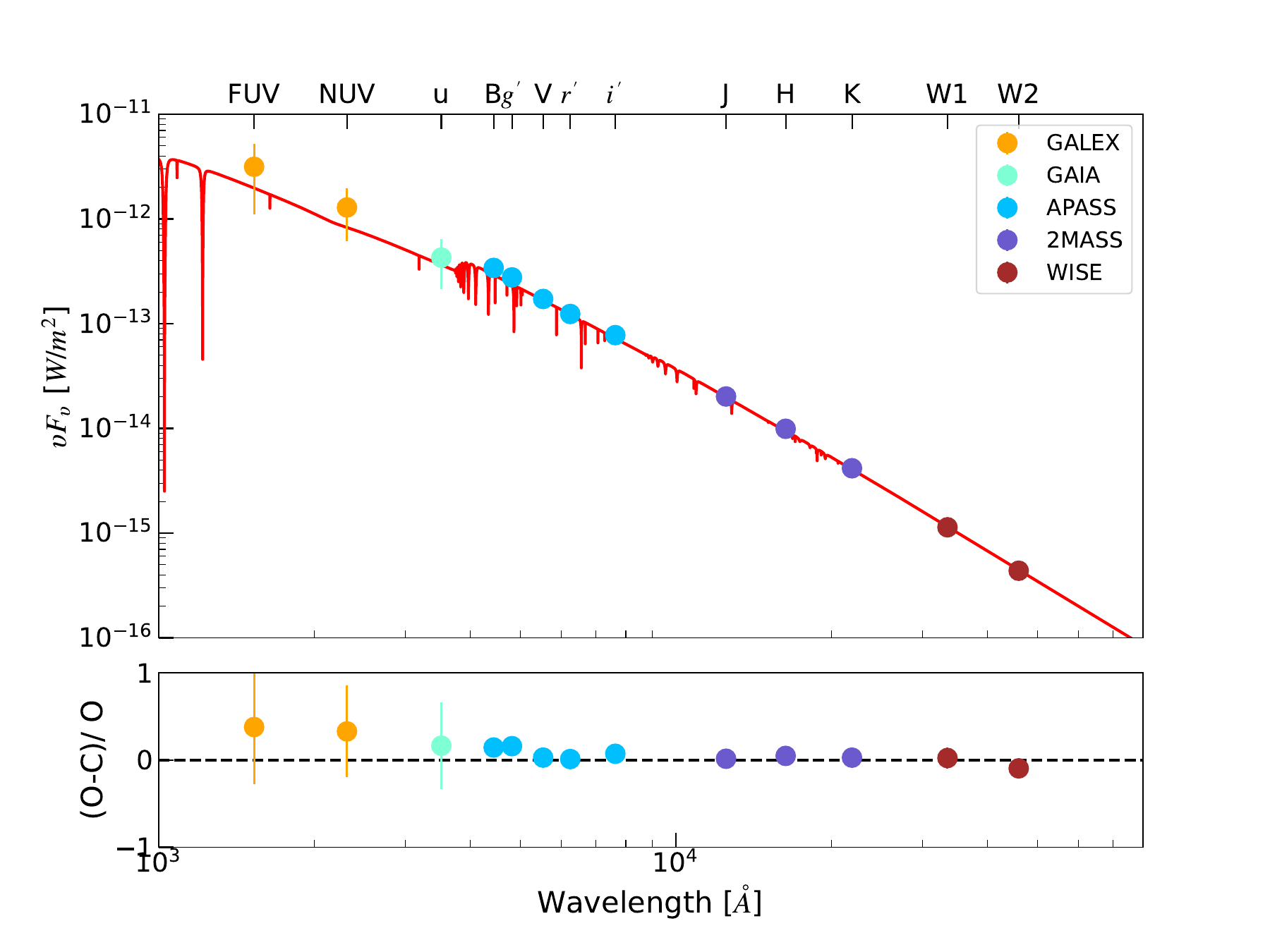} % 第二张图片
  \end{minipage}
  \hfill
  \begin{minipage}[b]{0.48\textwidth} % 创建一个minipage，宽度为页面宽度的一半
    \centering
    \includegraphics[width=\textwidth]{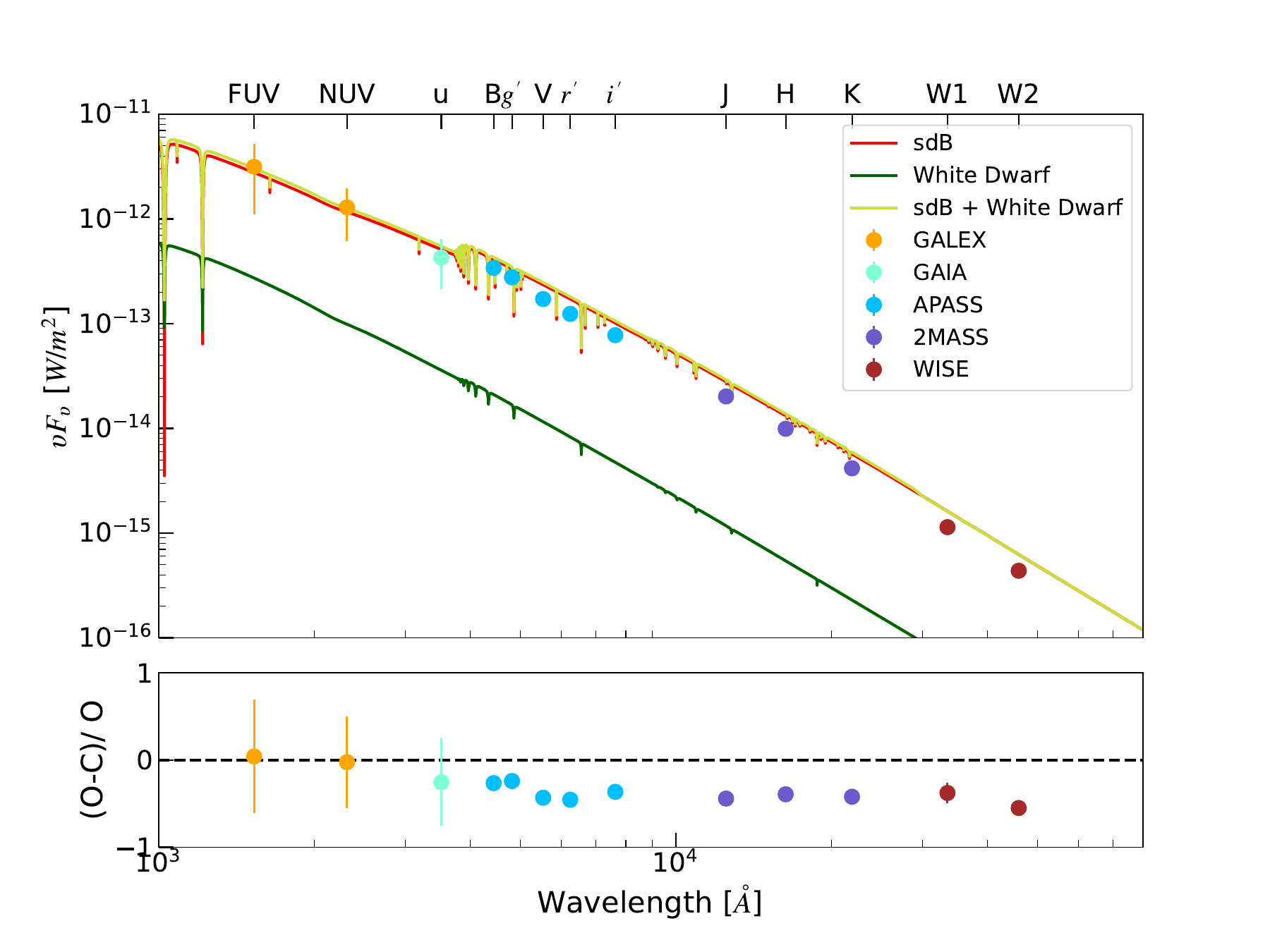} % 第三张图片
  \end{minipage}
  \caption{
SED fitting of J1710. 
The left panel displays the broad SED and observed photometry for the single sdB component (red), while the right panel includes an additional WD component (green). However, aligning the GALEX photometry values in the right panel results in a deviation of the total flux (yellow) from the observed photometry in other bands.
Photometric data sources include 
GALEX \citep{2008AJ....136..735L}, 
Gaia \citep[synthetic photometry derived from Gaia BP/RP mean spectra in SDSS u-band, aquamarine;][]{2022yCat.1360....0G}, 
APASS \citep{2015AAS...22533616H}, 
2MASS \citep{2003yCat.2246....0C}, and 
WISE \citep[W1 and W2, brown;][]{2014yCat.2328....0C}.
The Pan-STARRS (PS1) observations of J1710 in the $g$, $r$, $i$, $z$, and $y$ bands ($12.67$, $13.04$, $13.31$, $13.35$, and $13.62$ mag), obtained through the VizieR Photometry Viewer, exceeded the limiting magnitudes for bright stars in these bands as indicated in Table 11 of \citet{2016arXiv161205560C}, except for the $y$ band. Therefore, APASS photometric data were used instead of PS1 data.
}
  \label{fig:SED}
\end{figure*}

\subsection{Orbital parameters}
\label{sec:orbit}

Phase smearing correction was performed on the RV data from spectra, following the method outlined in \citet{2023AJ....165..119Y}. 
The corrected RVs were then analyzed using TheJoker \citep{2017ApJ...837...20P}, yielding the orbital parameters: orbital period \(P = 109.20279 \pm 0.00003\) minutes, eccentricity \(e = 0.013 \pm 0.011\), RV semi-amplitude \(K_1 = -223.3 \pm 3.8\) km/s, and systemic velocity \(V_0 = -40.5 \pm 7.3\) km/s. Corrected RV values are provided in Table \ref{tab.sp_obs}.

Considering J1710 as a compact binary system with a short period and a nearly circular orbit (\(e \approx 0\)), the orbit was assumed to be circular (\(e = 0\)). Using the aforementioned orbital parameters as priors and fixing the period at \(P = 109.20279\) minutes, the phase-folded RVs were fitted with a sinusoid, and Markov Chain Monte Carlo (MCMC) sampling was performed. The derived orbital parameters are \(K_1 = 222.2^{+2.7}_{-2.8}\) km/s, \(V_0 = -36.5 \pm 2.0\) km/s, and \(T_0 = 2459744.03573 \pm 0.00012\) days, as presented in Fig. \ref{fig:rvmcmc}.

Using the parameters from the best light-curve fit (see Section \ref{sec:wdfit}), we performed an MCMC analysis on the TESS light curve by varying both the period and \(T_0\). The light curve generated by the Wilson-Devinney code was expressed in phase and flux, while the observed TESS light curve was in time (dates) and flux. To compare the two, the observed light curve was folded into the phase domain for each combination of period and \(T_0\). We then used MCMC sampling to evaluate the match between the theoretical and observed light curves. The resulting ephemeris is:
\begin{multline}
T (\phi = 0) = 2459744.035491(25) \, \mathrm{BJD} \\
+ 0.07583527(5) \times E,
\end{multline}
where \textit{E} corresponds to the epoch, and \(\phi = 0\) marks the point when the visible star is farthest from the observer. 
Given the continuous and densely sampled TESS data from four sectors, the derived \(T_0\) is highly accurate, and this ephemeris is adopted as the final one.

When \(P = 109.20279\) minutes and \(K =222.2\) km/s, with \(e = 0\), the binary mass function is,
\begin{equation}
  f(M) = \frac{M_{1} \, q^3 \, \textrm{sin}^3 i} {(1+q)^{2}} = \frac{P \, K_{1}^{3} \, (1-e^2)^{3/2}}{2\pi G} = 0.086\, M_{\odot},
  \label{eqmass}
\end{equation}
where \(q = M_{2}/M_{1}\) denotes the mass ratio, and \(i\) signifies the orbital inclination.

\begin{figure}[htbp!]
%\figurenum{14}
\center
\includegraphics[width=0.45\textwidth]{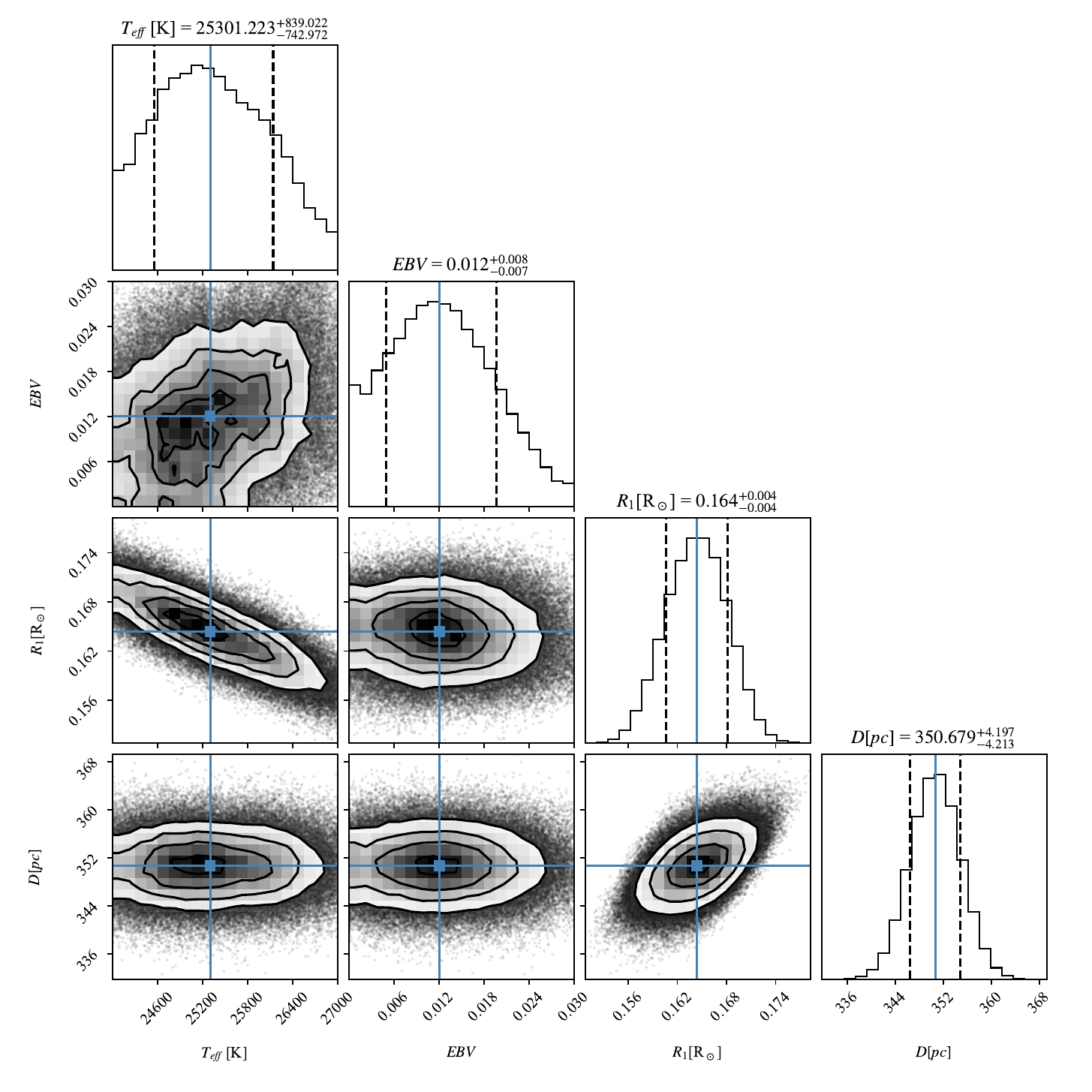}
\caption{
Corner plot of the SED fitting results for J1710 with a single component using the TMAP model.} 
%The posterior parameters are as follows: \(T_{\text{eff}} = 25301^{+839}_{-743} \, %\mathrm{K}\), \(E(B-V) = 0.012^{+0.008}_{-0.007}\), \(R_{1} = 0.164 \pm 0.004 \, R_{\odot}\), %and \(\text{D} = 350.68^{+4.20}_{-4.21} \, \mathrm{pc}\).

\label{fig:sed_mcmc}
\end{figure}

\subsection{Spectral energy distribution fitting}
\label{sec:sedfit}

\begin{figure*}[ht] % 使用figure*环境以跨越两列
  \centering
  \includegraphics[width=0.45\textwidth]{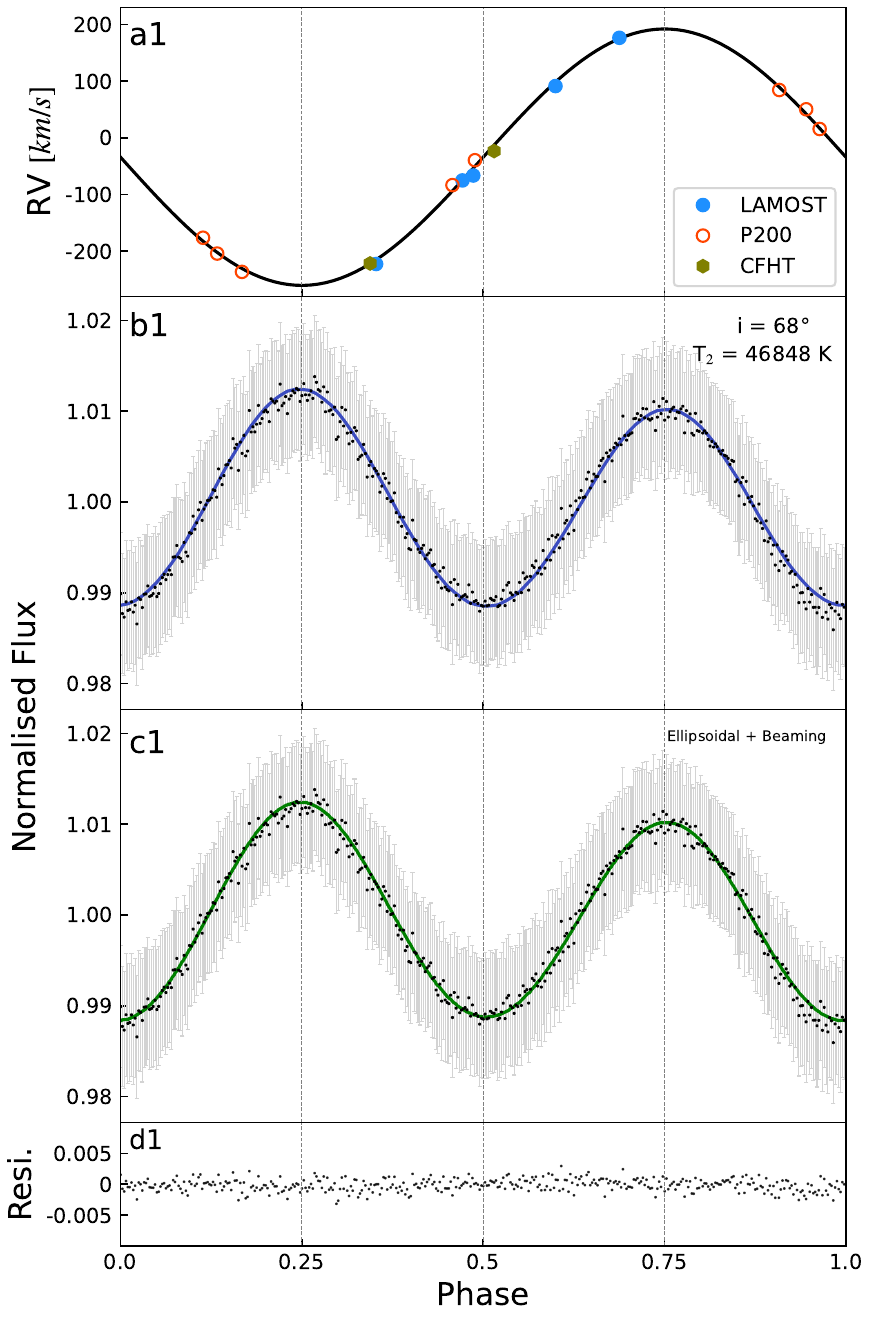} % 第一张图片
  \hfill
  \includegraphics[width=0.45\textwidth]{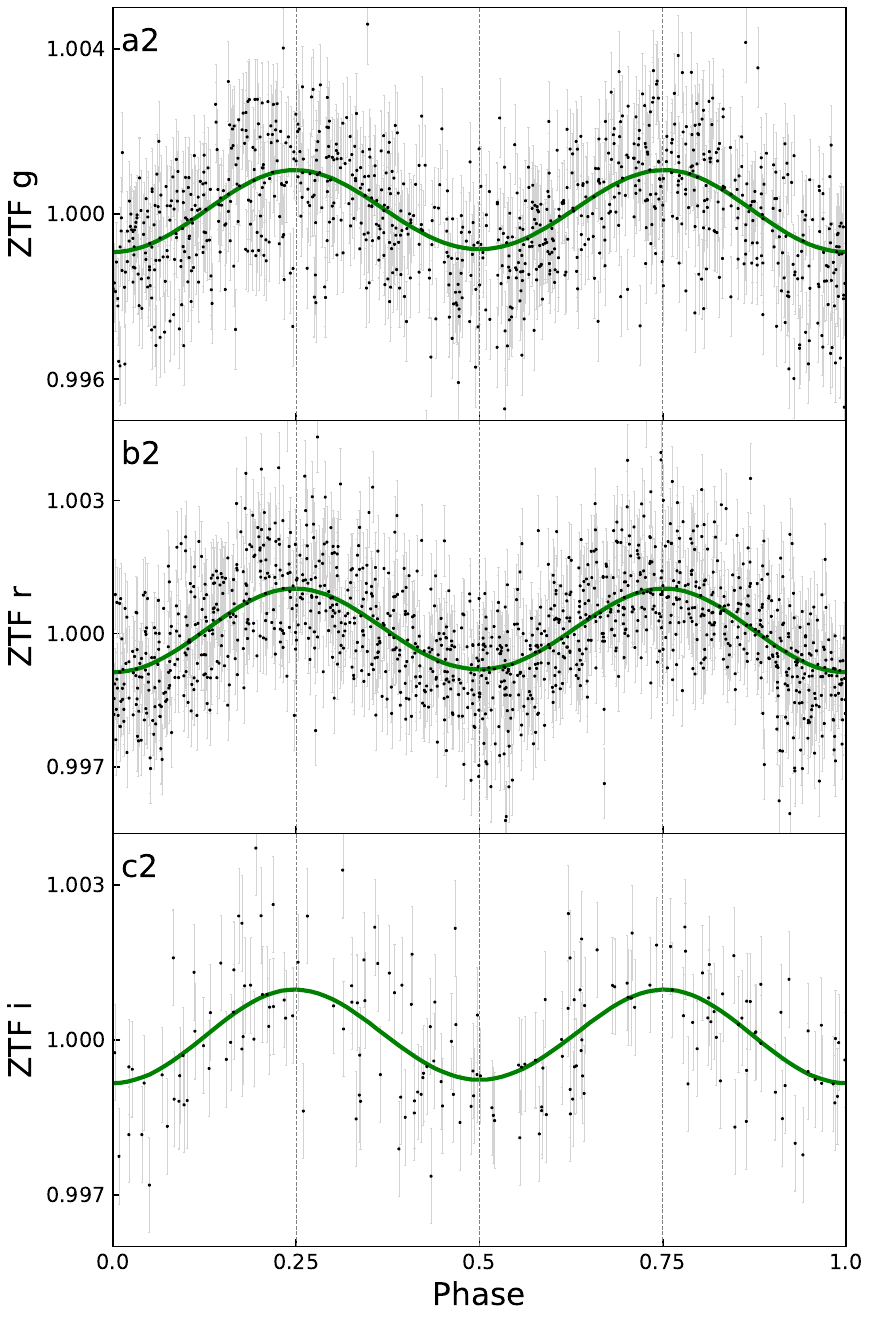} % 第二张图片
  \caption{
Folded RV curve, normalized light curves, and model light curves are presented in the ZTF \(g\), \(r\), and \(i\) bands, as well as the TESS band.
Photometric data points are represented as black points in all folded light curves, with gray shading indicating errors. 
The model light curves generated from the Wilson-Devinney code are depicted in green for their respective bands.
Panel a1 illustrates the calculated RV curve using the best-fit values \(K_1 =222.2 \, \text{km/s}\) and \(V_{\text{sys}} = -36.5 \, \text{km/s}\). Detailed spectroscopic information is provided in Table \ref{tab.sp_obs}.
Panel b1 displays the fitting results when the inclination is at its maximum and the companion star's temperature is at its highest, with the model light curve shown in blue. No significant features of the companion star are discernible.
Panel c1 presents the TESS light curve alongside the well-fitted model light curve, with the companion star's temperature fixed at 20000K and other parameters set to their mean values as described in the main text. The residuals are depicted in panel d1.
Panel a2, b2, and c2 show the ZTF light curves along with the corresponding model light curves.
}
  \label{fig:RVLC}
\end{figure*}

Correcting the Gaia DR3 \citep{2023A&A...674A...1G} parallax of J1710 to \(\varpi = 2.85 \pm 0.03\) mas according to \citet{2021A&A...649A...4L} yields a distance of \(350.87 \pm 3.70\) pc. Using this distance, the \texttt{Bayestar2019} map \citep{2019ApJ...887...93G} provides values of E(B-V) = 0.018, 0.018, and 0.027 for the $16th$, $50th$, and $84th$ percentiles of reddening at this coordinate, respectively. 
Additionally, from 2D dust maps by \citet{1998ApJ...500..525S} and \citet{2016A&A...596A.109P}, E(B-V) values of 0.013 and 0.02 are identified. 
Combining this information, E(B-V) = \(0.018 \pm 0.01\) is adopted and used as a prior for the spectral energy distribution (SED) fitting.

The SED of the visible star was fit using the models from the T\"uBingen NLTE Model-Atmosphere Package (TMAP) \citep{2012ascl.soft12015W}.
In the left panel of Fig. \ref{fig:SED}, photometric data spanning from the ultraviolet to the infrared were combined, and the SED was accurately fitted with a model of a single sdB (red line).

An MCMC approach is used to minimize the residuals between the observational SED and the TMAP model, enabling the derivation of errors on the parameters. We adopted prior assumptions for the effective temperature, surface gravity, and distance. The best-fitting parameters include the effective temperature \(T_{\rm{eff}} = 25301^{+839}_{-743} \, \mathrm{K}\), which is consistent with the value obtained from spectral fitting (\(T_{\rm{eff}} = 25164 \, \mathrm{K}\)), reddening \(E(B-V) = 0.012^{+0.008}_{-0.007}\), radius \(R_{1} = 0.164 \pm 0.004 \, R_{\odot}\), distance \(\rm{D} = 350.68^{+4.20}_{-4.21} \, \mathrm{pc}\), and luminosity \(\log (L/L_\odot) = 1.00 \pm 0.03\). The results of this analysis are presented in Fig. \ref{fig:sed_mcmc}, and these parameters are listed in Table \ref{tab.wdfit}.

Using the Koester Model \citep{2010MmSAI..81..921K} to account for the flux contribution from the companion WD, the result is shown in the right panel of Fig. \ref{fig:SED}. The sdB model flux remains consistent with that in the left panel. 
At a WD temperature of \(46848 \, \mathrm{K}\), the total model flux (yellow line) shows residuals close to zero in the ultraviolet band compared to the SED. However, significant deviations are observed in other bands, with the companion star contributing approximately 4\% of the optical bands' flux. 
Thus, the MCMC analysis likely captured only the parameters of the visible star, indicating that the companion WD's effective temperature should be lower than \(46848 \, \mathrm{K}\). Consequently, the companion WD's luminosity contribution can be neglected.

\subsection{Extinction and kinematics }
\label{sec:extinction}

Based on the extinction law of \citet{1989ApJ...345..245C} and the effective wavelengths of various bands, when considering $E(B-V) = 0.012^{+0.008}_{-0.007}$, the extinctions calculated in the $g$, $r$, $i$, Gaia $G$, and TESS bands are listed in Table \ref{tab.wdfit}. The mean apparent magnitude and absolute magnitude in the Gaia $G$ band are 12.59 and $4.81 \pm 0.04$, respectively.

The proper motion for J1710 provided by Gaia DR3 is \(\text{pmra} = 0.75 \pm 0.05 \, \text{mas/yr}\) and \(\text{pmdec} = -13.34 \pm 0.05 \, \text{mas/yr}\). 
Additionally, the system velocity is \(-36.5 \pm 2.0\,\text{km/s}\), resulting in a 3D motion in LSR of \(-28.6 \pm 0.3\,\text{km/s}\), \(-20.5 \pm 1.2\,\text{km/s}\), \(-14.9 \pm 0.9\,\text{km/s}\), indicating a thin disk object (\(P_{\text{thin}} \approx 98.3\%\)) \citep{1987AJ.....93..864J, 2013ApJ...764...78R}. 
Here, the bootstrapping method was employed to estimate the uncertainties.

\subsection{Light curve fitting}
\label{sec:wdfit}

\begin{figure*}[ht]
%\figurenum{14}
\center
\includegraphics[width=0.98\textwidth]{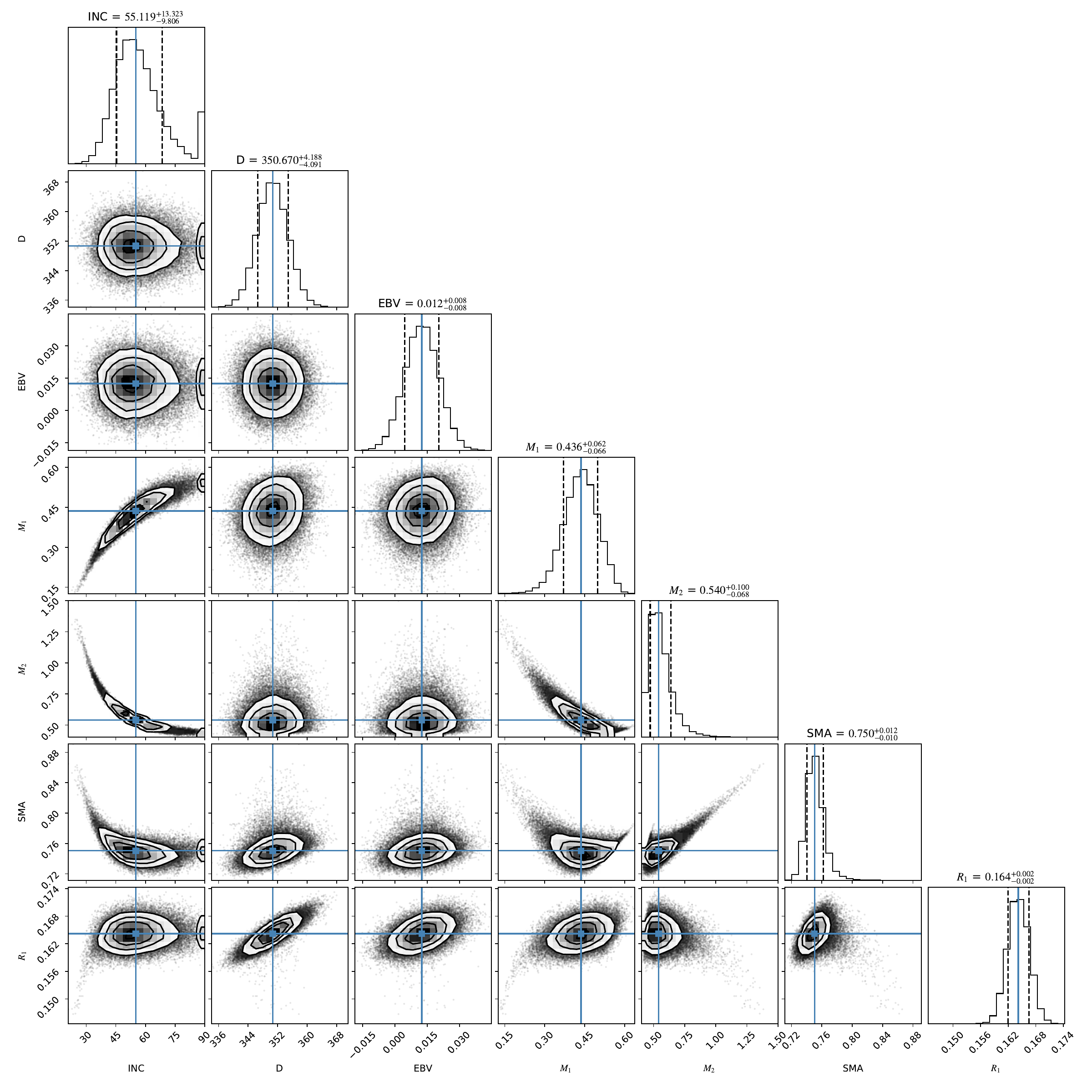}
\caption{
Bootstrapping sampling results for the TESS band light curve using the Wilson-Devinney code. 
The diagonal elements present the probability distribution of each parameter, with percentiles of 16\%, 50\%, and 84\% denoted by vertical lines. Note that star 1 denotes the visible component, while star 2 denotes the invisible compact component. The uncertainties in inclination, distance, extinction, gravity darkening, and other parameters were all considered. See the text for more details.
}
\label{fig:bootstrap}
\end{figure*}

Using the established ephemeris, the TESS data were phase-folded and re-binned into 400 data points, while the ZTF data were phase-folded for each band. The resulting normalized light curves are shown in Fig. \ref{fig:RVLC}, where the primary star's ellipsoidal variation dominates. Additionally, Doppler beaming is observed, leading to an asymmetric light curve due to the large RV amplitude, which results in higher flux when the visible star is approaching Earth compared to when it is moving away \citep{2012ApJ...749...42H}.

To determine the binary's properties, the Wilson-Devinney code \citep{1971ApJ...166..605W, 1979ApJ...234.1054W, 1990ApJ...356..613W, 2012AJ....144...73W} was used to fit both the TESS light curves and RVs simultaneously. Assuming a circular orbit and synchronized stellar rotation periods with the orbital period, the projected rotational velocity \(v_{\text{rot}} \sin i = 89 \pm 12 \, \mathrm{km/s}\), corresponding to an orbital inclination \(i = {55^{+13}_{-10}}^{\circ}\). In this compact binary system composed of an sdB and a WD, the reflection effect is negligible due to the much smaller size of the WD. 

Some additional information was incorporated as input for the orbital solution. The orbital period was fixed at 109.20279 minutes, as derived in Section \ref{sec:orbit}. The primary star's temperature was constrained to \(25,301\, \mathrm{K}\), as determined in Sect. \ref{sec:sedfit}. Additionally, the primary star's radius, obtained from high-precision SED fitting and Gaia parallax data, served as a reliable reference.
Limb darkening coefficients in the logarithmic form (\(e = 0.27\), \(f = 0.15\)) and Doppler beaming coefficients (\(\beta = 1.47\)) were fixed based on the tables of \citet{2020A&A...634A..93C}, with values closest to the atmospheric parameters for the TESS filter. Given the high effective temperature of the primary star, the gravity darkening exponent (\(\mathrm{GR}\)) and bolometric albedo (\(\mathrm{ALB}\)) were set to 1. The companion star, with no visible observational features, had its temperature fixed at 20,000 \(\mathrm{K}\), and a large potential \(\Omega_{2}\) was assigned to ensure a small radius.

Through 50,000 bootstrap fitting samples, the uncertainty of the orbital solution was evaluated, and the final results are presented in Fig. \ref{fig:bootstrap}. 
The mass ratio \(q\), semi-major axis (SMA), orbital inclination \(i\), and the gravitational potential of the visible star \(\Omega_{1}\) were treated as free parameters. 
The orbital inclination \(i = {55^{+13}_{-10}}^{\circ}\), distance \(D = 350.68 \pm 4.20\) pc, and extinction \(A_T = 0.022 \pm 0.014\) were sampled using Gaussian distributions. 
For each set of sampled parameters, the code iterated to derive optimal values for the semi-major axis SMA, mass ratio q, and gravitational potential \(\Omega_{1}\), along with the corresponding inclination \(i\).
The derived values for \(M_{1}\), \(M_{2}\), \(R_{1}\), and \(\log g_{1}\) are \(M_{1} = 0.44^{+0.06}_{-0.07}\, M_{\odot}\), \(M_{2} = 0.54^{+0.10}_{-0.07}\, M_{\odot}\), \(R_{1} = 0.164 \pm 0.002\, R_{\odot}\), \(SMA = 0.75 \pm 0.01\, R_{\odot}\), \(q = 1.24^{+0.47}_{-0.28}\), and \(\log g_{1} = 5.65^{+0.05}_{-0.07}\) dex, with the optimal zero-point flux in the TESS-band \(\mathrm{CALIB}\) of \(0.1734 \, \mathrm{erg\,s^{-1}\,cm^{-3}}\).
The sdB star fills approximately 61\% of its Roche lobe.
Due to the absence of eclipses in the light curve, larger uncertainties remain for parameters such as mass. The best-fitting model is shown in panel c1 of Fig. \ref{fig:RVLC}.

%Since the ZTF magnitudes are calibrated using the Pan-STARRS system \citep{2019PASP..131a8003M}, J1710 exceeds the observation limit in the $g$, $r$, and $i$ bands of the Pan-STARRS system \citep{2016arXiv161205560C}, rendering the magnitudes provided by ZTF unreliable. 
%Therefore, the TESS light curve was exclusively fitted.

To assess the impact of the companion star's temperature and orbital inclination on the light curve fitting, the orbital inclination was fixed at its maximum value, \(i = 68^\circ\), while the temperature of the companion star was adjusted to its maximum value, \(T_2 = 46848\, \mathrm{K}\), as depicted in panel b1 of Fig. \ref{fig:RVLC}. Under these conditions, the model light curve (blue line) remains consistent with the TESS data (black dots) without showing distinctive features. Consequently, this scenario does not provide additional constraints on parameters such as the temperature of the companion star. Thus, as discussed in Sect. \ref{sec:sedfit}, the flux contribution from the companion star could be disregarded. For all other cases, the temperature of the companion star was held constant at \(T_2 = 20\,000\, \mathrm{K}\).
With the inclination fixed at \(55^\circ\) and considering the effects of ellipsoidal variation and Doppler beaming, as shown in panel c1, the model light curve (green line) fits well with the TESS data, with residuals depicted in panel d1.

According to \citet{1983ApJ...268..368E}, the Roche radius of the companion can be approximated as
\begin{equation}
R_{\text{RL}} = \frac{0.49 \, q^{2/3} \, a}{0.6 \, q^{2/3} + \ln(1+q^{1/3})},
\label{Rrl}
\end{equation}
where \(a\) is the orbital separation between the two stars. Kepler's third law relates the orbital separation and period. 
For \(M_2\) ranging from \(0.47\) to \(0.64\, M_{\odot}\), the Roche radius \(R_{\text{RL2}}\) varies from \(0.28\) to \(0.32\, R_{\odot}\), all of which are smaller than the radii of main-sequence stars with the corresponding mass.
The absence of emission lines in the spectra suggests that J1710 does not exhibit strong mass transfer, indicating that the companion must be a WD with a radius smaller than its Roche radius.

J1710 is saturated in the PS1 images in the \(g\), \(r\), and \(i\) bands and falls within the ZTF saturation limits of approximately 12.5 to 13.2 magnitudes\footnote{\url{https://irsa.ipac.caltech.edu/data/ZTF/docs/ztf_extended_cautionary_notes.pdf}} \citep{2016arXiv161205560C, 2019PASP..131a8003M}. Moreover, the ZTF magnitudes exhibit significant scatter, making the data unreliable for light curve fitting.
Therefore, using the best-fit parameters from the light curve model, synthetic light curves were generated in the \(g\), \(r\), and \(i\) bands, normalized, and compared with the observed ZTF light curves. Since the ZTF magnitudes are calibrated using the Pan-STARRS system \citep{2019PASP..131a8003M}, the ZTF magnitudes were converted to SDSS using the PS1-to-SDSS relation \citep{2012ApJ...750...99T}, as shown in panels a2, b2, and c2 of Fig. \ref{fig:RVLC}.
The amplitudes of J1710's light curves are influenced by various factors, including the temperature and ellipticity of the primary star, the relative sizes of the stars, and surface brightness.
The model light curves in these three bands (green lines) generally agree with the ZTF data (black dots, with gray lines representing errors), thereby validating the reliability of our results.

\section{Discussions}
\label{sec.mesa}

\subsection{Modelling the sdB}
\label{sec:mesasdb}

\begin{figure*}[htbp!]
\center
\includegraphics[width=0.98\textwidth]{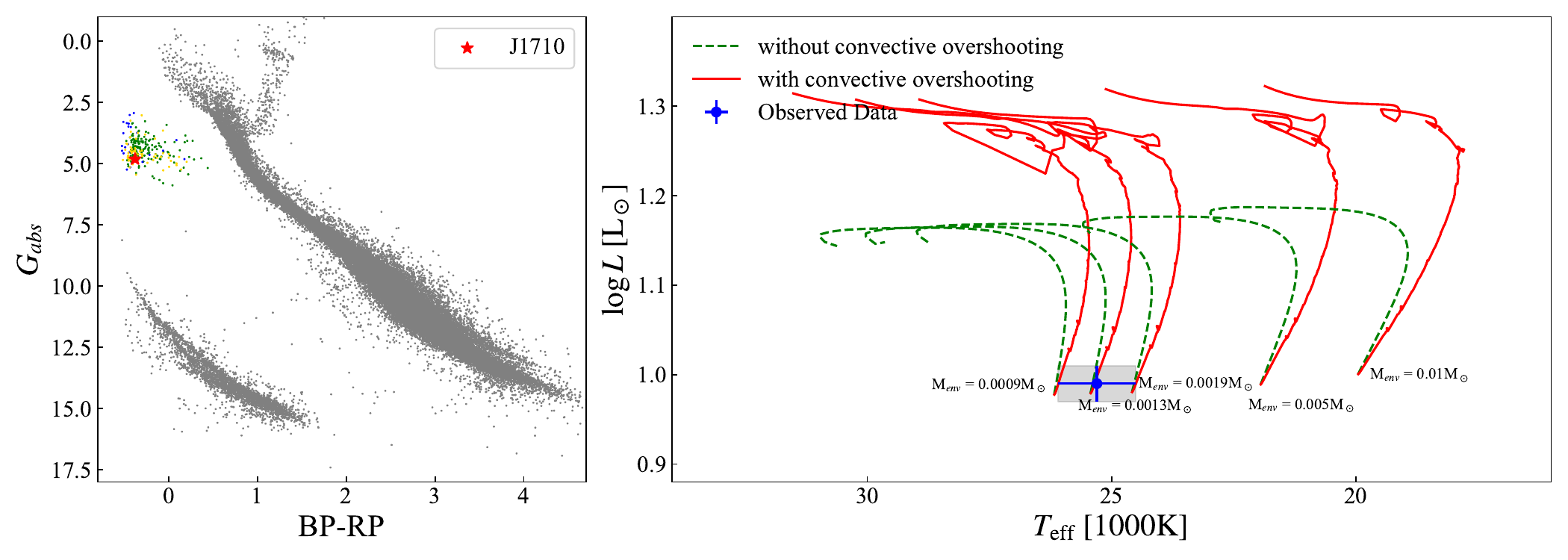}
\caption{
Color-magnitude diagram and evolutionary tracks for five sdB models derived from a \(2.16\,M_{\odot}\) main-sequence progenitor. 
In the left panel, gray data points represent sources within 100 parsecs in the Gaia DR3 dataset, while green, blue, and gold points indicate candidate sources identified as sdB, sdO, and sdOB stars, respectively.
Right panel shows MESA evolutionary tracks of sdB model stars with different hydrogen envelope masses, with or without convective overshooting. When the helium core mass is 0.431 \(M_{\odot}\) and the envelope mass is 0.0013 \(M_{\odot}\), the evolutionary track covers the observed data and corresponding 1-\(\sigma\) error region.
}
\label{fig.MESAandHR}
\end{figure*}

The system's position on the color-magnitude diagram is depicted in the left panel of Fig. \ref{fig.MESAandHR}. 
Different samples are represented by different colors: green for sdB, blue for sdO, and gold for sdOB, following the definitions provided in \citet{2021yCat..18810135L}, with respective sample sizes of 169, 22, and 63. 
A sample of objects within a 100 parsec radius is depicted as gray points, consistent with the approach used in \citet{2019MNRAS.488.2892P}, where they are shown as black points in their Figure 1. 
Comparison with Figure 1 of \citet{2016PASP..128h2001H} further emphasizes that the visible star in J1710 is a sdB.

\begin{table*}
\centering
\caption{Summaries of parameters for J1710.%Parameters below the dashed line are output results.
\label{tab.wdfit}}
%\end{minipage}
\setlength{\tabcolsep}{4.5pt}
\begin{center}
\begin{tabular}{cccc}
\hline\noalign{\smallskip}
Name & Description & Unit & Value \\
\hline\noalign{\smallskip}
RA & Right ascension J2000 & [degrees] & 257.5550469 \\
DEC & Declination J2000 & [degrees] & +53.4461208 \\
P & Orbital period & [days] & 0.07583527(5) \\
\(T_0\) & Barycentric Julian Day & [days] & 2459744.035491(25) \\
\(V_0\) & System velocity & [km/s] & $-36.5 \pm 2.0$ \\
\(K_1\) & RV semi-amplitude & [km/s] & $222.2^{+2.7}_{-2.8}$ \\
\hline\noalign{\smallskip}
Plx & Parallax & [mas] & $2.85\pm0.03$ \\
D & Distance & [pc] & $350.68^{+4.20}_{-4.21}$ \\
E(B-V) & Reddening in B-V color & [mag] & $0.012^{+0.008}_{-0.007}$ \\
Ag & Extinction in g band & [mag] & 0.045$\pm$0.028 \\
Ar & Extinction in r band & [mag] & 0.033$\pm$0.020 \\
Ai & Extinction in i band & [mag] & 0.025$\pm$0.016 \\
AT & Extinction in TESS band & [mag] & 0.022$\pm$0.014 \\
AG & Extinction in Gaia G band & [mag] & 0.049$\pm$0.018 \\
ALB & Bolometric albedos & [-] & $1.0$ \\
GR & Gravity darkening exponent & [-] & $1.0$ \\
CALIB & TESS band zero point & [$\mathrm{erg\,s^{-1}\,cm^{-3}}$] & 0.1734 \\
\hline\noalign{\smallskip}
$T_{\rm eff}$ & Spectral temperature & [K] & $25164\pm663$ \\
log $g$ & Spectral surface gravity & [dex] & 5.68$\pm$0.08 \\
\(\log (n(\text{He})/n(\text{H}))\) & Spectral metallicity & [dex] & -2.26$\pm$0.09 \\
\(v_{rot} \sin i\) & Projected rotational velocity & [km/s] & 89$\pm$12 \\
$T_{\rm eff}$ & SED temperature & [K] & $25301^{+839}_{-743}$ \\
\(R_1\) & SED star 1 radius & [$R_{\odot}$] & $0.164\pm0.004$ \\
\(L_1\) & SED star 1 luminosity & [$\log (L/L_\odot)$]  & \(1.00 \pm 0.03 \) \\
\hline\noalign{\smallskip}
\(M_1\) & Light curve primary mass & [$M_{\odot}$] & $0.44^{+0.06}_{-0.07}$ \\
\(R_1\) & Light curve primary radius & [$R_{\odot}$] & $0.164\pm0.002$ \\
log $g$ & Light curve surface gravity & [dex] & $5.65^{+0.05}_{-0.07}$ \\
SMA & Semi major axis & [$R_{\odot}$] & 0.75$\pm$0.01 \\
%INC & Orbital Inclination & [degrees] & $54-90$ \\
\(M_2\) & Light curve secondary mass & [$M_{\odot}$] & $0.54^{+0.10}_{-0.07}$ \\
\hline\noalign{\smallskip}
$T_{\rm eff}$ & MESA sdB temperature & [K] & $25284$ \\
log $g$ & MESA sdB surface gravity & [dex] & $5.64$ \\
\(M_1\) & MESA sdB mass & [$M_{\odot}$] & $0.432$ \\
\(R_1\) & MESA sdB radius & [$R_{\odot}$] & $0.164$ \\
% R1 & MESA primary radius & [$R_{\odot}$] & $0.164$ \\
\hline\noalign{\smallskip}
\end{tabular}
\end{center}
\smallskip
\end{table*}

The results obtained so far suggest that J1710 comprises a compact binary system housing an sdB with a period shorter than two hours. The visible star is estimated to have a mass of \(M_{1}=0.44^{+0.06}_{-0.07}\, M_{\odot}\), while the companion star, a WD, is estimated to possess a mass of \(M_{2}=0.54^{+0.10}_{-0.07}\, M_{\odot}\). According to the classical formation theory of sdBs, the primary energy loss mechanism involves the ejection of the sdB's hydrogen envelope through a CE phase \citep{2002MNRAS.336..449H, 2003MNRAS.341..669H, 2016PASP..128h2001H}. This process leads to a significant reduction in the orbital period.

To understand the current evolutionary status of the sdB, we used the code MESA, specifically version r23.05.1 \citep{2011ApJS..192....3P, 2013ApJS..208....4P, 2015ApJS..220...15P, 2018ApJS..234...34P, 2019ApJS..243...10P, 2023ApJS..265...15J}, to model single-star evolution. 
Models were initialized with a metallicity \(Z=0.02\) and initial masses ranging from 1 to 5 \(M_\odot\), starting from the pre-main sequence phase. For simplicity, atomic diffusion mixing, stellar wind, and rotation were excluded. At the tip of the red giant branch, the built-in tool named 'Relax Mass' was utilized to artificially remove the hydrogen envelope, aiming to determine the helium core mass and hydrogen envelope mass at the birth of the sdB. This procedure aimed to match the current observational parameters of the visible star (\(T_{\text{eff}} = 25301^{+839}_{-743}\, K\), \(\log (L/L_\odot) = 1.00 \pm 0.03\)).

When modeling the evolution of sdBs and considering convective overshooting, applying exponential overshoot with the recommended value from \citet{2000A&A...360..952H} (\(f_{\text{ov}} = 0.016\)) at all convective boundaries, with the mixing length parameter \(\alpha\) set to 1.8, results in breathing pulsations when the core helium abundance (\(Y_c\)) is less than 0.2. These pulsations can cause fluctuations in the core size with stellar luminosity and radius \citep{1989A&A...221...27C, 2021ApJ...923..166L}. Despite the long recognition of breathing pulsations, uncertainties persist, and there is a possibility that they may be numerical artifacts rather than genuine physical phenomena \citep{2019ApJS..243...10P}. 
Therefore, for reference, sdBs were also modeled without overshooting. The stellar evolution calculations were terminated when helium was depleted at the stellar center (\(Y_c < 10^{-3}\)).

The right panel of Fig. \ref{fig.MESAandHR} depicts that, for an initial stellar mass of \(2.16 \, M_{\odot}\), the sdB model—comprising a \(0.431 \, M_{\odot}\) helium core and a \(0.0013 \, M_{\odot}\) hydrogen envelope—closely matches the observed parameters of the visible star (indicated by the blue cross). 
Naturally, this result is based on the specific MESA version and the physical parameter settings used in our analysis. For other typical sdB models, such as one with a mass of \(0.47 \, M_{\odot}\), the temperature, luminosity, and \(\log g\) values do not correspond to the observed data presented in Sect. \ref{sec:results}. 
The sdB parameters derived from our model are \(T_{\text{eff}} = 25 284 \, \mathrm{K}\), \(\log (L/L_{\odot}) = 1\), \(\log g = 5.64\) dex, and \(R_1 = 0.164 \, R_{\odot}\), with an age of approximately 12 Myr in the sdB phase. 
These parameters show strong consistency with the observed results in Sect. \ref{sec:results}, thereby mutually affirming the robustness and reliability of the obtained results.

\subsection{Binary evolution}
\label{sec:mesabinary}
We employed the MESA binary evolution module to evolve the sdB and WD in J1710 together, investigating when the sdB will fill its Roche lobe, and whether J1710 will evolve into an AM CVn system or, alternatively, whether GW radiation will dominate orbital shrinkage, leading to a DWD system merger. Assuming fully conservative mass transfer from the sdB to the WD during the RLOF phase, we followed the mass transfer rates prescribed by \citet{1988A&A...202...93R}. 
All angular momentum losses were assumed to arise from gravitational wave radiation.

\begin{figure}[htbp!]
\centering
\includegraphics[width=0.48\textwidth]{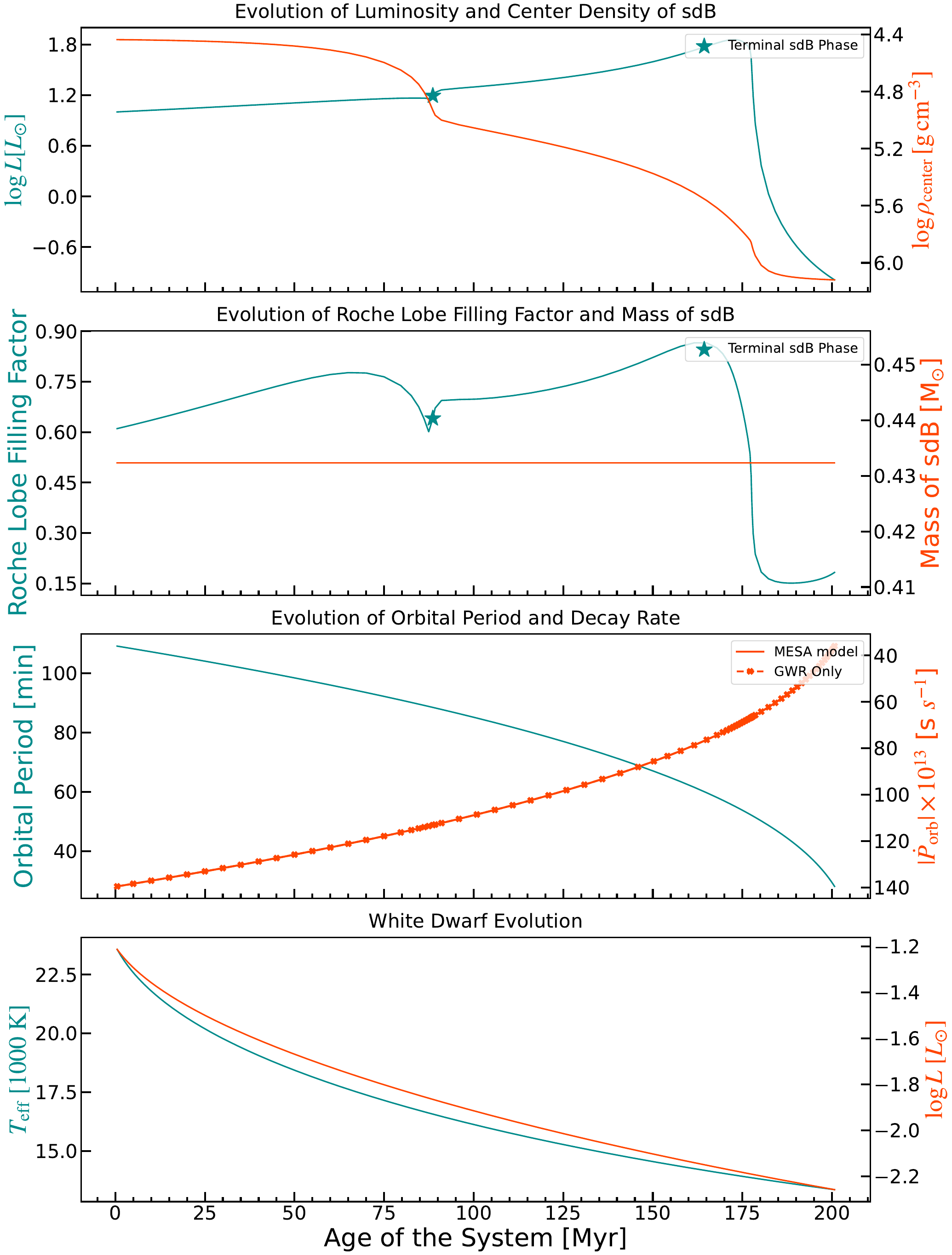}
\caption{
Binary evolution of J1710 from the present day until the sdB star evolves into a WD, modeled using the MESA code. In the top two panels, the moment when the sdB phase ends is marked with a star symbol. In each subplot, the green lines correspond to the left vertical axis, and the red lines correspond to the right vertical axis. Detailed descriptions of the panels are provided in Sect. \ref{sec:mesabinary}.
}
\label{fig.MESABinary}
\end{figure}

The current sdB models were obtained as described in Sect. \ref{sec:mesasdb}, with evolutionary tracks stopping when they reached the observed $T_{\mathrm{eff}}$ and $\log(L/L_\odot)$ values. Due to the lack of detailed information about the WD, we adopted the results from the SED fit in Sect. \ref{sec:sedfit}, which suggest that the WD temperature should be below 46848 K. Assuming an sdB mass of $0.432\, M_\odot$, the WD mass was determined to be $0.54\, M_\odot$ based on the mass function. We constructed a WD model with a C/O core using the \texttt{make\_co\_wd} test case in MESA, and evolved a $0.54\, M_\odot$ WD model until $\log(L/L_\odot) = -1.2$, where the WD temperature is approximately 23701 K, consistent with the SED fit. 
The initial orbital period was set to 109.20279 minutes, and the evolution was halted when the luminosity of the primary (sdB) dropped below $\log(L/L_\odot) = -1$. The final results, shown in Fig. \ref{fig.MESABinary}, indicate that the J1710 system will evolve into a DWD and eventually merge.

Fig. \ref{fig.MESABinary} illustrates the binary system's evolution from the present until the sdB becomes a WD. The top panel shows the evolution of the primary star's luminosity (green line) and central density (red line) over time. Approximately 89 Myr from now, when the core helium abundance of the sdB drops to $Y_c < 10^{-3}$ (marked with a star), the sdB phase ends and the star transitions to the WD phase. When the primary's luminosity falls below $\log(L/L_\odot) = -1$ and its central density exceeds $6\, \text{g\,cm}^{-3}$, the sdB has fully transformed into a WD. 

The second panel displays the Roche lobe filling factor (green line) and the mass of the sdB (red line) over time. The Roche lobe filling factor, defined as the ratio of the sdB’s radius to its Roche lobe radius, remains below 1 throughout the evolution, indicating that no significant mass transfer occurs. Moreover, the sdB's mass remains constant, further confirming the absence of mass exchange before the system becomes a DWD. Thus, no AM CVn system is formed.

The third panel shows the orbital period (green line) and its decay (red lines) over time. The red lines compare the orbital period decay predicted by the MESA model with that expected from GW radiation alone. The lines overlap, indicating that orbital shrinkage and angular momentum loss are solely driven by GW emission. The fourth panel illustrates the decreasing temperature (green line) and luminosity (red line) of the WD as it cools over time. The results demonstrate that the sdB star will not fill its Roche lobe before it becomes a WD, and no mass transfer occurs in the system.

When the primary star has a mass of $M_1 = 0.432\, M_{\odot}$ and the secondary has a mass of $M_{2} = 0.54^{+0.10}_{-0.07}\, M_{\odot}$, GW radiation is the only mechanism responsible for orbital shrinkage, ultimately leading to the merger of the two degenerate companions. The merger timescale for such a double degenerate system can be estimated using the following formula \citep{2021ApJ...921..160C}:
\begin{equation}
    \tau_{\text{GW}} = 10 \left(\frac{M_1 + M_2}{M_1 M_2}\right)^{1/3} P_{\text{hours}}^{8/3}\, \text{Myr}.
    \label{mergerT}
\end{equation}
With an orbital period of 1.82 hours, the merger timescale ranges from 180 to 231 Myr, roughly matching the evolutionary age of the system indicated in Fig. \ref{fig.MESABinary}, confirming that the J1710 binary system will merge within a Hubble time.

\subsection{In the absence of tidal synchronization}
\label{sec:tidal}

Tidal synchronization is a debated topic for stars with convective cores and radiative envelopes like sdBs, as the exact dissipation mechanisms remain elusive. In Sect. \ref{sec:wdfit}, we assumed tidal synchronization. The tidal synchronization time can be estimated simply as \citep{2001icbs.book.....H}:
\begin{equation}
t_{\text{sync}} = 10^4 \left[\frac{(1+q)}{2q}\right]^2 P^4,
  \label{sync}
\end{equation}
where \(t_{\text{sync}}\) is in years. Assuming an orbital eccentricity \(e = 0\), \(q = M_2/M_1 = 1.24\) as derived in Sect. \ref{sec:wdfit}, and \(P = 109.20279\) minutes, the synchronization time is about 0.27 years. Therefore, J1710 likely reaches tidal synchronization very quickly.

However, current observations suggest that tidal synchronization has not been achieved in many sdB binaries. For example, in the J162256+473051 system, the orbital period is 0.069789 days, while the rotation period is 0.115156 days, indicating that the sdB star's rotation is not synchronized with its orbital motion \citep{2014A&A...564A..98S}. Similar non-synchronous behavior is observed in other systems, such as KIC 11179657 \citep{2012MNRAS.422.1343P} and PG1142-037 \citep{2016MNRAS.458.1417R}, which also display longer rotation periods than expected for synchronization. 
These observations further reinforce the notion that tidal synchronization is not a common feature in many sdB binary systems.

Moreover, \citet{2018MNRAS.481..715P} estimated the tidal synchronization timescale using tidal dissipation theories that account for dissipation via turbulent convection. Their results suggest that in most observed cases, the synchronization timescale significantly exceeds the lifetime of the sdB star. 
However, recent work by \citet{2024arXiv240816158M} challenges these assumptions, particularly in regard to the treatment of tidal dissipation through gravity waves. While earlier models, such as those by \citet{2010A&A...519A..25G}, \citet{2012MNRAS.422.1343P}, and \citet{2018MNRAS.481..715P}, assumed efficient damping of gravity waves near the stellar surface, \citet{2024arXiv240816158M} suggest that these waves may be less effectively damped than previously thought, potentially leading to much faster tidal synchronization. Their findings indicate that the synchronization timescale is primarily determined by the orbital period, with less dependence on the mass of the companion star.

Although the tidal synchronization assumption in Sect. \ref{sec:wdfit} yields an sdB mass consistent with the MESA models from Sect. \ref{sec:mesasdb}, we cannot entirely rule out the possibility that J1710 is not tidally synchronized due to the lack of additional observational data. 
In the absence of tidal synchronization, the sdB star’s rotational period would be expected to exceed the orbital period, allowing for larger orbital inclinations compared to when tidal synchronization is present. Re-fitting the TESS light curve and RV data simultaneously using the Wilson-Devinney code, and permitting the orbital inclination \(i\) to vary between \(45^\circ\) and \(90^\circ\), we find an inclination of \(i = 67^\circ \pm 15^\circ\), with 1-\(\sigma\) uncertainties yielding \(M_1 = 0.49^{+0.06}_{-0.07}\, M_\odot\), \(M_2 = 0.48^{+0.08}_{-0.03}\, M_\odot\), \(R_1 = 0.164 \pm 0.002\, R_\odot\), \(SMA = 0.75 \pm 0.01\, R_\odot\), \(q = 0.97^{+0.38}_{-0.14}\), and \(\log g_1 = 5.69^{+0.05}_{-0.06}\) dex. These results remain consistent with the MESA model predictions within the error margins.

Thus, while the absence of tidal synchronization introduces additional uncertainties in determining the orbital parameters, it does not alter the conclusion that J1710 is a compact sdB+WD binary system with an orbital period of less than 2 hours.

\subsection{Gravitational waves}
\label{sec:gw}
Considering the maximum possible companion star mass \(M_2 = 0.64 \, M_{\odot}\) for the J1710 binary system, with \(M_1 = 0.432 \, M_{\odot}\), and \(P = 109.20279\) minutes, the dimensionless GW amplitude (\(\mathcal{A}\)) is calculated from \citet{2012A&A...544A.153S} as:
\begin{equation}
\mathcal{A} = \frac{2(GM_c)^{5/3}}{c^4d} \left(\pi f_{\text{gw}}\right)^{2/3},
\label{eq:A}
\end{equation}
where the GW frequency \(f_{\text{gw}} = 2/P\), and the distance \(d = 350.68 \, \text{pc}\). This yields \(f_{\text{gw}} \approx 0.3 \, \text{mHz}\) and dimensionless GW amplitude \(\mathcal{A} \approx 2 \times 10^{-22}\).

The characteristic strain amplitude \(h_c\) of J1710 based on 4 years of LISA observations can be calculated using the formula \citep{2015CQGra..32a5014M}:
\begin{equation}
h_c = \sqrt{f_{\text{gw}} T_{\text{obs}}} \mathcal{A},
\label{eq:hc}
\end{equation}
where \(T_{\text{obs}} = 4\) years. This results in a characteristic strain \(h_c \approx 4 \times 10^{-20}\).
Unfortunately, this falls notably short of the sensitivity threshold outlined in the LISA sensitivity curve (refer to Figure 3 in \citet{2018MNRAS.480..302K} and Figure 5 in \citet{2021ApJ...921..160C}). Nonetheless, the GW signal emitted by J1710 will introduce additional foreground noise, exerting an influence on LISA's sensitivity curve \citep{2023LRR....26....2A}. We anticipate that future space-based GW observatories will detect more short-period WD binaries and their merger events, enhancing our understanding of DWD evolution.

\section{Conclusion}
\label{sec.conclusion}
This paper provides a comprehensive analysis of the nearby compact binary system J1710, comprising an sdB and a WD. 
The system features a short orbital period of 109.20279(7) minutes, less than 2 hours, and is located at a distance of \(350.68^{+4.20}_{-4.21} \, \mathrm{pc}\) from Earth. Through spectroscopic, SED, and light curve fitting, the visible star is identified as an sdB with a temperature \(T_{\text{eff}} = 25301^{+839}_{-743} \, \mathrm{K}\), radius \(R_1 = 0.164 \pm 0.002 \, R_{\odot}\), and luminosity \(\log (L/L_\odot) = 1 \pm 0.03\). 
The visible star's mass is determined to be \(M_1 = 0.44^{+0.06}_{-0.07} \, M_{\odot}\) through light curve fitting, while the MESA model covering the observed temperature and luminosity indicates \(M_1 = 0.432\, M_{\odot}\). The system's membership in the thin-disk binary population is suggested by dynamic analysis.

In the absence of direct evidence for the companion star, its mass varies with the orbital inclination. As indicated in Sect. \ref{sec:wdfit}, the companion star's mass is estimated to be \(M_2 = 0.54^{+0.10}_{-0.07} \, M_{\odot}\), and it is identified as a WD.
As described in Sect. \ref{intro.sec}, only six detached systems with a WD companion and \(P_{\text{orb}} < 2\,\text{hr}\) have undergone comprehensive analysis, including J1710. Excluding CD-\(30^{\circ}1122\), which is at a distance of \(349.65^{+8.77}_{-8.35} \, \mathrm{pc}\) \citep{2024MNRAS.527.2072D}, J1710 is the closest at \(350.68^{+4.20}_{-4.21} \, \mathrm{pc}\). Furthermore, its Gaia G-band mean magnitude stands at 12.59, indicating exceptional brightness and making it a prime candidate for additional observations. The abundance of observational data could potentially augment our comprehension of the condition and evolution of such compact binary systems.

High-resolution ultraviolet and optical observations of J1710 with high S/N could enhance our comprehension of its evolutionary trajectory and unveil distinctive post-CE phase characteristics. It is anticipated to undergo evolution into a DWD system and eventually merge, thus emerging as a pivotal source of low-frequency GWs for space-based observatories.

\begin{acknowledgements}
This work is supported by the National Natural Science Foundation of China (12273056, 12090041, 11933004) and the National Key R\&D Program of China (2019YFA0405002, 2022YFA1603002).
HL.Y and ZR.B are supported by the National Key R\&D Program of China (Grant No. 2023YFA1607901).
HL.Y acknowledges support from the Youth Innovation Promotion Association of the CAS (Id. 20200060) and National Natural Science Foundation of China (Grant No. 11873066).

ZW.L and XF.C are supported by the National Natural Science Foundation of China (NSFC, Nos. 12288102, 12473034, 12125303, 12090040/3), the National Key R\&D Program of China (Nos. 2021YFA1600403/1 and 2021YFA1600400), the Yunnan Fundamental Research Projects (Nos. 202401AT070139, 202201AU070234, 202101AU070276), the Natural Science Foundation of Yunnan Province (Nos. 202201BC070003, 202001AW070007), the International Centre of Supernovae, Yunnan Key Laboratory (No. 202302AN360001), and the "Yunnan Revitalization Talent Support Program"-Science and Technology Champion Project (No. 202305AB350003).
We also acknowledge the science research grant from the China Manned Space Project with No.CMS-CSST-2021-A10.

Guoshoujing Telescope (the Large Sky Area Multi-Object Fiber Spectroscopic Telescope LAMOST) is a National Major Scientific Project built by the Chinese Academy of Sciences. Funding for the project has been provided by the National Development and Reform Commission. LAMOST is operated and managed by the National Astronomical Observatories, Chinese Academy of Sciences. 

This work presents results from the European Space Agency (ESA) space mission Gaia. Gaia data are being processed by the Gaia Data Processing and Analysis Consortium (DPAC). Funding for the DPAC is provided by national institutions, in particular the institutions participating in the Gaia MultiLateral Agreement (MLA). The Gaia mission website is https://www.cosmos.esa.int/gaia. The Gaia archive website is https://archives.esac.esa.int/gaia. 
This work uses data obtained through the Telescope Access Program (TAP), which has been funded by the TAP member institutes. We would like to acknowledge and thank TAP (ID: CTAP2022-A0018) for their support.
We acknowledge use of the VizieR catalog access tool, operated at CDS, Strasbourg, France, and of Astropy, a community-developed core Python package for Astronomy (Astropy Collaboration, 2013). 
\end{acknowledgements}


\bibliographystyle{aa.bst} % style aa.bst
\bibliography{aa.bib} % your references Yourfile.bib

\end{CJK*}
\end{document}